\documentclass[11pt]{article}
\pdfoutput=1
\usepackage{geometry}             
\usepackage{graphicx}
\usepackage{amssymb}
\usepackage{amsmath}
\usepackage{epstopdf}
\usepackage{comment}
\usepackage{cite}

\usepackage[hyperindex=true,
          pdfstartview=FitH,
          bookmarksnumbered=true,
          bookmarksopen=true,
          citecolor=blue,
          linkcolor=blue,
          colorlinks=true,
          unicode]{hyperref}

\newcommand{\be}{\begin{equation}}
\newcommand{\ee}{\end{equation}}
\newcommand{\bea}{\begin{eqnarray}}
\newcommand{\eea}{\end{eqnarray}}

\renewcommand{\d}[1]{\ensuremath{\operatorname{d}\!{#1}}}

\begin{document}
\title{Black Hole Evaporation in Ho\v{r}ava-Lifshitz Gravity}
\author{Hao Xu , Yen Chin Ong\\
Center for Gravitation and Cosmology, College of Physical Science and Technology, \\Yangzhou University, 180 Siwangting Road, Yangzhou City, Jiangsu Province 225002, China\\
{\em email}:
\href{mailto:haoxu\_phys@163.com}{haoxu\_phys@163.com},
\href{mailto:ycong@yzu.edu.cn}{ycong@yzu.edu.cn}\\}

\author{Hao Xu \thanks{{\em
        email}: \href{mailto:haoxu\_phys@163.com}
        {haoxu\_phys@163.com}}, Yen Chin Ong\thanks{{\em
        email}: \href{mailto:ycong@yzu.edu.cn}
        {ycong@yzu.edu.cn}}\\
Center for Gravitation and Cosmology, \\College of Physical Science and Technology, Yangzhou University, \\180 Siwangting Road, Yangzhou city, Jiangsu Province 225002, China\\}

\date{}                             
\maketitle

\begin{abstract}

Ho\v{r}ava-Lifshitz (HL) gravity was formulated in hope of solving the non-renormalization problem in Einstein gravity and the ghost problem in higher derivative gravity theories by violating Lorentz invariance. In this work we consider the spherically symmetric neutral AdS black hole evaporation process in HL gravity in various spacetime dimensions $d$, and with detailed balance violation parameter $0\leqslant \epsilon^2\leqslant 1$. We find that the lifetime of the black holes under Hawking evaporation is dimensional dependent, with $d=4,5$ behave differently from $d\geqslant 6$. For the case of $\epsilon=0$, in $d=4,5$, the black hole admits zero temperature state, and the lifetime of the black hole is always infinite. This phenomenon obeys the third law of black hole thermodynamics, and implies that the black holes become an effective remnant towards the end of the evaporation. As $d\geqslant 6$, however, the lifetime of black hole does not diverge with any initial black hole mass, and it is bounded by a time of the order of $\ell^{d-1}$, similar to the case of  Schwarzschild-AdS in Einstein gravity (which corresponds to $\epsilon^2=1$), though for the latter this holds for all $d\geqslant 4$.
The case of $0<\epsilon^2<1$ is also qualitatively similar with $\epsilon=0$.

\end{abstract}

\section{Introduction}

Black holes radiate once quantum mechanics is taken into account \cite{Hawking:1974rv,Hawking:1974sw}. Hawking radiation now serves as a tool to study the relations between different areas of physics, including general relativity, quantum field theory, quantum statistics, and quantum information.
Due to the Hawking radiation, if there is no incoming matter to balance the energy loss, the black hole would keep losing mass and thermal entropy until it evaporates away. This raises the problem about the information loss paradox \cite{Almheiri:2012rt,Hawking:2016sgy}. For a static uncharged black hole with initial mass $M_0$ in four-dimensional asymptotically flat spacetime in Einstein gravity, it has a lifetime $t\sim M_0^3$ \cite{Page:1976df}. Similar analysis can also be carried out for rotating \cite{Page:1976ki} or charged black holes \cite{Hiscock:1990ex,Xu:2019wak}. For the rotating black hole, it will lose angular momentum several times faster than the mass, thus any initially rotating black hole will evolve into a static Schwarzschild case (unless there are a lot of scalar particle species \cite{9710013, 9801044}). Compared to the lifetime of a Schwarzschild black hole with the same initial mass, the presence of angular momentum can increase the lifetime by at most a factor of about 3. For the charged black hole, although both mass and charge are decreasing functions of time and the black hole evolves toward a Schwarzschild state eventually, some charged black holes can increase their charge-over-mass ratio very near to the extremal Reissner-Nordstr\"om limit during the evaporation process \cite{Hiscock:1990ex,Xu:2019wak}, and the lifetime of these black hole may be extended by a huge factor, compared to the lifetime of a neutral black hole (rotating or not) with the same initial mass.

On the other hand, features of the evaporation process of black holes with nontrivial asymptotic behavior can be very different from their counterparts in the case of asymptotically flat spacetime. For a spherical black hole in asymptotically AdS spacetime, with the usual reflective boundary condition imposed, Hawking radiation from a sufficiently large black hole can be re-absorbed and eventually thermal equilibrium is achieved. However, if an absorbing AdS boundary condition is chosen so that the radiation particles cannot reflect back \cite{Avis:1977yn}, the lifetime of black hole in $d$-dimensional spacetime does not diverge even if the initial black hole mass is taken to be arbitrarily large. Using geometrical optics approximation we know it is bounded by a time of the order $\ell^{d-1}$, where $\ell$ is the AdS curvature radius \cite{Page:2015rxa}.

However, all of the above analysis on the black hole evolution is carried out in the context of Einstein gravity, which possesses a dimensional gravitational coupling constant in mass units $[G_N]=-2$. It is well known that, when we apply the quantum field theory to Einstein gravity in an attempt to obtain a quantum gravity theory, we will be facing a serious resistance: Einstein gravity, at the quantum loop level, is not a (perturbatively) renormalizable theory \cite{tHooft:1974toh,Goroff:1985th,Kiefer2012,Weinberg1980}. In other words, from the point of view of quantum field theory, Einstein gravity is only an effective theory which will break down in the high energy regime. Power-counting analysis shows that in order to have a renormalizable theory, the dimension of  gravitational coupling constant should be larger than or equal to zero, otherwise the perturbative effective quantum field theory will break down at high energy, which is exactly what happens in Einstein gravity.

One way to have an improved ultraviolet (UV) behavior is to introduce higher order derivative corrections to the Einstein-Hilbert action \cite{Stelle:1976gc}, such as the quadratic term $R_{\mu\nu}R^{\mu\nu}$ used in conformal gravity \cite{Maldacena:2011mk} and Lovelock gravity \cite{Lovelock:1971yv}, so that the gravitational propagator is modified at high energy and UV divergence can be cured. However, this modification in turn causes the field equations to have time derivatives whose order is greater than two. For example, the field equations are fourth-orders in the quadratic case. This will introduce a ghost that makes the theory non-unitary. In addition, according to the theorem of Ostrogradsky \cite{Woodard:2015zca} , such a system is not kinematically stable, thus the higher derivative gravity theory with Lorentz invariance is problematic.

Another possible way out of this problem is to violate the Lorentz invariance in the UV, so that we can have high-order spatial derivative terms in the Lagrangian, while keeping the time derivative terms to the second order. This model, known as Ho\v{r}ava-Lifshitz (HL) gravity, which was proposed by Petr Ho\v{r}ava in 2009 \cite{Horava:2009uw}, suggested that there is an anisotropy between time and space at short distance, which is measured by a Lifshitz-type dynamical critical exponent $z$. Essentially, HL gravity describes a non-relativistic renormalizable gravity theory at short distance, and the UV behavior is power-counting renormalizable at large energy scale. Lorentz invariance, on the other hand, is presented as an accidental symmetry at large distance, and the classical Einstein gravity is restored in the IR limit.

It should be emphasized that, usually, Lorentz invariance is taken as one of the fundamental principles in physics. It is also strongly supported by observations \cite{Mattingly:2005re,Liberati:2013xla}, and so far there is no conclusive evidence that indicates such a symmetry can be broken at high enough energy scales (there are, however, some suggestive signs from high energy astrophysics \cite{1810.03571, 1810.01652,1906.07329}). However, according to our current understanding on the quantum gravity theory, space and time are quantized at the Planck energy scale, and continuous spacetime only emerges as a classical limit at sufficiently low energy scale. Since Lorentz invariance is derived from the continuous symmetry of spacetime, it is reasonable to assume that it can be broken in the UV. This is what Ho\v{r}ava did in his original work, and this idea has been extensively developed in the past 10 years. Related work includes, but is not restricted to, the self-consistency of the theory, cosmology, black hole and thermodynamics, non-relativistic gauge/gravity duality, quantization of HL gravity, etc. See e.g. \cite{Mattingly:2005re,Mukohyama:2010xz,Sotiriou:2010wn,Horava:2011gd,Wang:2017brl} for related reviews.

There is a caveat here: due to the violation of Lorentz invariance in HL gravity, ``black holes'' are rather tricky to be defined, as the horizon radii generically depend on the energies of test particles \cite{1105.4259}. In fact, once Lorentz invariance is broken, particles can travel as arbitrarily large speed. In the words of \cite{0910.5487v1}, ``it is not clear when and for whom they are black''. It was later discovered that there \emph{is} still a causal boundary in HL gravity, which was called the ``universal horizon'' \cite{1404.3413,1408.5976,1501.04134}, a concept first introduced in the context of khrononmetric theory of gravity \cite{1110.2195}. For an asymptotically flat Schwarzschild black hole, the corresponding ``universal horizon'' is the hypersurface $r=3M/2$, instead of the event horizon $r=2M$ \cite{1501.04134}. The universal horizons exhibit thermodynamical properties \cite{1202.4497,1210.4940,1309.0907}, and should be thought as being associated with the Hawking temperature in HL gravity and Lorentz-violating theories in general \cite{1312.0405, 1512.01900, 1703.08207}, much like the event horizon is associated with Hawking radiation in general relativity. Nevertheless, in the geometric optics limit, the \emph{effective emission surface} is not the event horizon even in general relativity. Therefore, in our analysis below, this subtlety of universal horizon vs. event horizon does not arise. There is of course the issue that whether it is the temperature at the universal horizon that should be considered instead of the standard Hawking one associated the event horizon. Leaving this technicality aside, in this work we will employ the latter. In any case, this probably would not affect the qualitative pictures -- which is what we are concerned with -- by much. See e.g.\cite{Cognola:2016gjy,Casalino:2018wnc} for more discussion on the instability problems in HL gravity.

The first black hole solution in HL gravity, which describes a spherically symmetric black hole in four dimensional AdS spacetime with a general dynamical coupling constant $\lambda$ \cite{Lu:2009em} , was discovered by Lu, Mei and Pope soon after Ho\v{r}ava's original paper, followed shortly by solutions of topological black holes \cite{Cai:2009pe}. Similar analysis was also extended to cases with various critical exponent $z$, spacetime dimensions $d$, and detailed balance violation parameter $\epsilon$ \cite{Li:2014fsa}. In fact, even in the limit $\lambda=1$, in which the asymptotically AdS solutions can be obtained in the large distance approximation, these black holes also have very different behaviors from their counterparts in Einstein gravity. Therefore, it is important to study the various properties of HL black holes. In this work, we
shall consider the black hole evaporation process in HL gravity and compare the results with the AdS black hole in Einstein gravity \cite{Page:2015rxa} or higher derivative gravity theories \cite{Xu:2017ahm,Xu:2018liy,Xu:2019krv}.

Specifically, in the present work we investigate the spherically symmetric neutral AdS black hole evaporation process in HL gravity, focusing on the $\lambda=1$ and $z=3$ with various spacetime dimensions $d$ and detailed balance violation parameter $0\leqslant \epsilon^2\leqslant 1$. Just as \cite{Page:2015rxa}, we consider the case in which one starts in AdS with a large black hole, whose horizon radius is much larger than the AdS length scale $\ell$. This black hole solution and its thermodynamics in HL gravity have been investigated in \cite{Li:2014fsa}. In the next section we will present a brief review of the black hole solution and its thermodynamics. In Sec.(\ref{(III)}), we investigate the black hole evaporation process in different dimensions $d$ and different values of $\epsilon$. In the final section we give some concluding remarks. We adapt the natural unit system, setting the speed of light in vacuum $c$, the gravitational constant $G_N$, the Planck constant $h$ and the Boltzmann constant $k$ equal to one.

\section{Thermodynamics of Ho\v{r}ava-Lifshitz Gravity}

For the $d$-dimensional spacetime in HL gravity at the $z=3$ UV fixed point, the gravitational action can be written as \cite{Li:2014fsa}
\begin{align}
  S&=\int\mathrm{d}t\bigg[L_0+(1-\epsilon^2)L_1\bigg],\label{action}\\
  L_0&\equiv\int\mathrm{d}^{d-1}x\sqrt{g}N\bigg[\frac{2}{\kappa^2}\bigg(K_{ij}K^{ij}-\lambda\,K^2\bigg)
  +\frac{\kappa^2}{8\kappa^4_{W}}\frac{\Lambda_{W}}{1-(d-1)\lambda}\bigg((d-3)R-(d-1)\Lambda_{W}\bigg) \bigg],\nonumber\\
  L_1&\equiv\int\mathrm{d}^{d-1}x\sqrt{g}N\frac{\kappa^2}{8\kappa^4_{W}}\frac{1}{1-(d-1)\lambda}
  \bigg[\left(1-\frac{d-1}{4}-\lambda\right)R^2
  -(1-(d-1)\lambda)R^{ij}R_{ij} \bigg].\nonumber
\end{align}
The first two terms in $L_0$ are the kinetic actions, while the others correspond to the potential actions. $R_{ij}$ and $R$ are the Ricci tensor and Ricci scalar respectively. $K_{ij}=\frac{1}{2N}\bigg(\dot{g}_{ij}-\nabla_{i}N_{j}-\nabla_{j}N_{i}\bigg)$ is based on the ADM decomposition of the higher dimensional spacetime, i.e. $\mathrm{d}s^2_{d}=-N^2\mathrm{d}t^2+g_{ij}(\mathrm{d}x^i-N^i\mathrm{d}t)(\mathrm{d}x^j-N^j\mathrm{d}t)$. The lapse, shift and $(d-1)$-metric $N$, $N^i$ and $g_{ij}$ are all functions of $t$ and $x^i$, and the dot denotes a derivative with respect to $t$. There are five constant parameters in the action: $\Lambda_{W}$, $\lambda$, $\epsilon$, $\kappa$ and $\kappa_{W}$, corresponding to the cosmological constant, dynamical coupling constant, detailed balance violation parameter, gravitational constant, and the speed of light respectively. We will focus on the case in which $\lambda=1$ and $\Lambda_{W}<0$, where the Einstein gravity in AdS spacetime can be obtained in large distance approximation. The $\epsilon$ parameter satisfies $0\leqslant \epsilon^2\leqslant 1$, where $\epsilon=0$ corresponds to the detailed-balance condition and $\epsilon=1$ recovers Einstein gravity.

The spherically symmetric neutral black hole can be described by the metric \cite{Li:2014fsa}
\begin{equation}
\d s^2 = -f(r)\d t^2 + \frac{\d r^2}{f(r)} +r^{2}\d{\Omega}^2_{d-2},
\end{equation}
where $\d {\Omega}^2_{d-2}$ is the round metric of the $(d-2)$-sphere, and the function $f(r)$ takes the form
\begin{align}
  f(r)=1-\frac{2\Lambda_{W}}{(1-\epsilon^2)}\frac{r^2}{(d-2)(d-3)}
  -r^{\frac{5-d}{2}}\sqrt{\frac{c_0}{(1-\epsilon^2)}+\frac{\epsilon^2}{(1-\epsilon^2)^2}\frac{4\Lambda_{W}^2r^{d-1}}{(d-2)^2(d-3)^2}}.
\end{align}
Here $c_0$ is an integration constant related to the black hole mass by
\begin{align}
  M=-\frac{\Omega_{d-2}}{16\pi}\frac{1}{(d-3)\Lambda_{W}}c_0=-\frac{c_0}{\Lambda_{W}},
\end{align}
where we have chosen the natural units and without loss of generality we set $\Omega_{d-2}=16(d-3)\pi$. The black hole event horizon $r_+$ is defined as the largest root of $f(r)=0$ in AdS spacetime. In our case, for $d=4$ there are two roots of $f(r)=0$, which corresponds to the black hole event horizon and an inner horizon. For $d\geq5$, there is only one root corresponding to the event horizon. We can write the black hole mass as the function of $r_+$. This gives
\begin{align}
M=-\frac{4\Lambda_{W}r_+^{d-1}}{(d-3)^2(d-2)^2}+\frac{4r_+^{d-3}}{(d-2)(d-3)}-\frac{(1-\epsilon^2)r_+^{d-5}}{\Lambda_{W}}.
\label{mass}
\end{align}
The temperature is proportional to the surface gravity of the black hole horizon
\begin{align}
T=\frac{4(d-1)\Lambda_{W}^2r_+^4-4(d-2)(d-3)^2\Lambda_{W}r_+^2+(1-\epsilon^2)(d-2)^2(d-3)^2(d-5)}{8\pi (d-2)r_+(-2\Lambda_{W}(d-3)r_+^2+(1-\epsilon^2)(d-2)(d-3)^2)}.
\label{temperature}
\end{align}
The other thermodynamical quantities, such as the entropy and specific heat, can also be evaluated. We shall omit them here.

Since we are considering the system in AdS spacetime with a negative cosmological constant $\Lambda_{W}$, from Eq.\eqref{mass} we know that the black hole mass is always positive. It is a monotonic function of $r_+$ and it goes to zero as $r_+\rightarrow 0$. However, it is worth emphasizing that the features of the black hole temperature \eqref{temperature} in $r_+\rightarrow 0$ are determined by the last term $(1-\epsilon^2)(d-2)^2(d-3)^2(d-5)$ in the numerator.

\begin{figure}[h!]
\begin{center}
\includegraphics[width=0.32\textwidth]{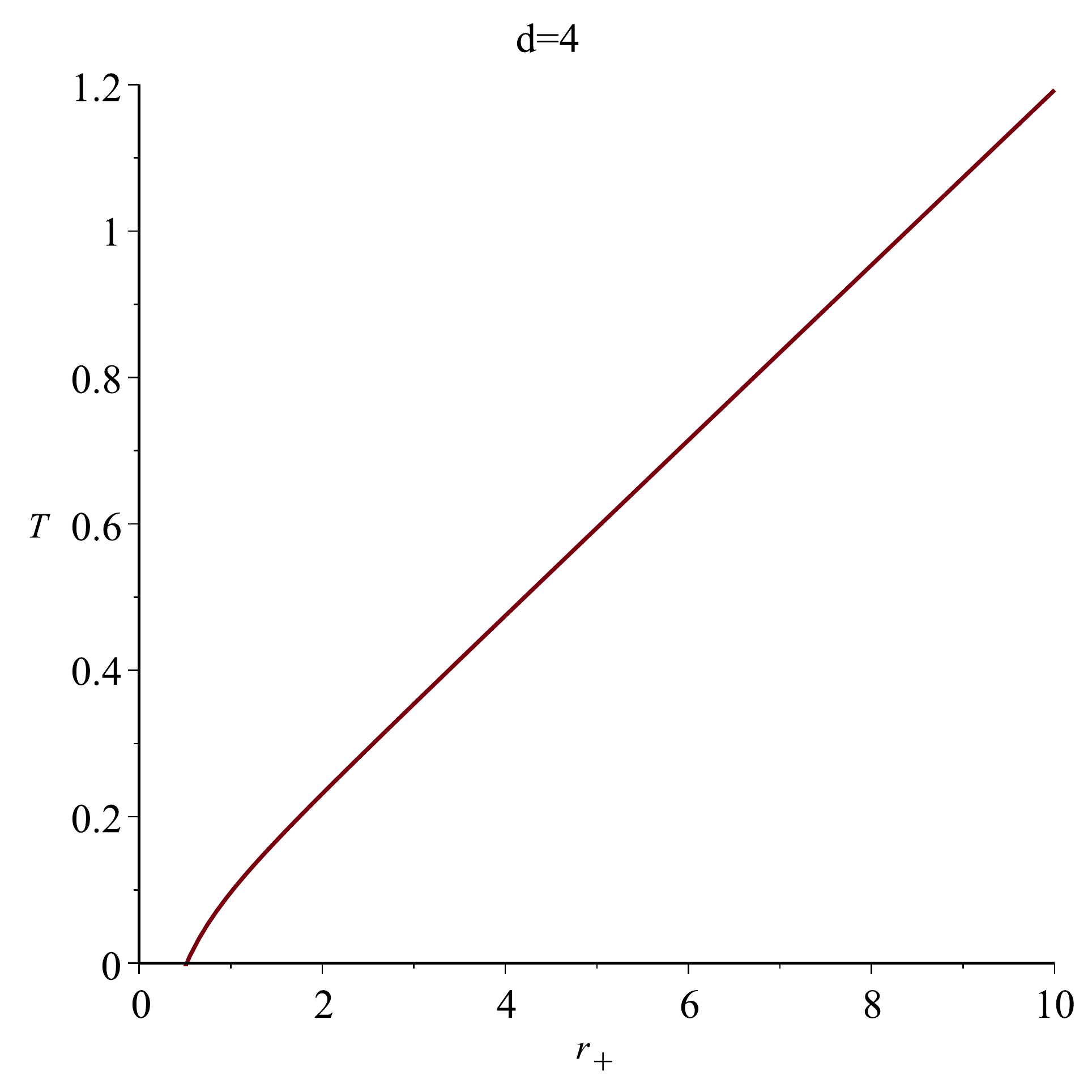}
\includegraphics[width=0.32\textwidth]{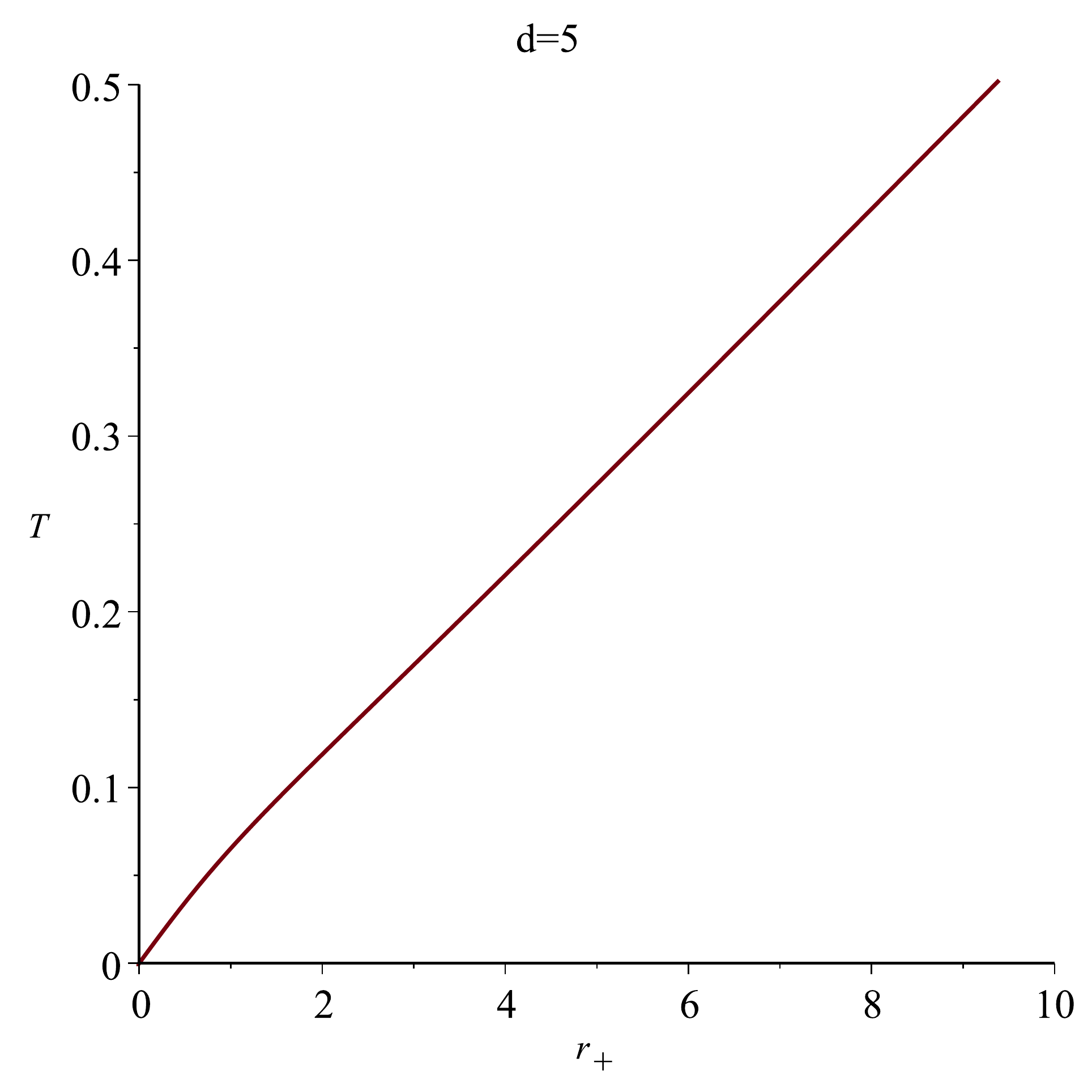}
\includegraphics[width=0.32\textwidth]{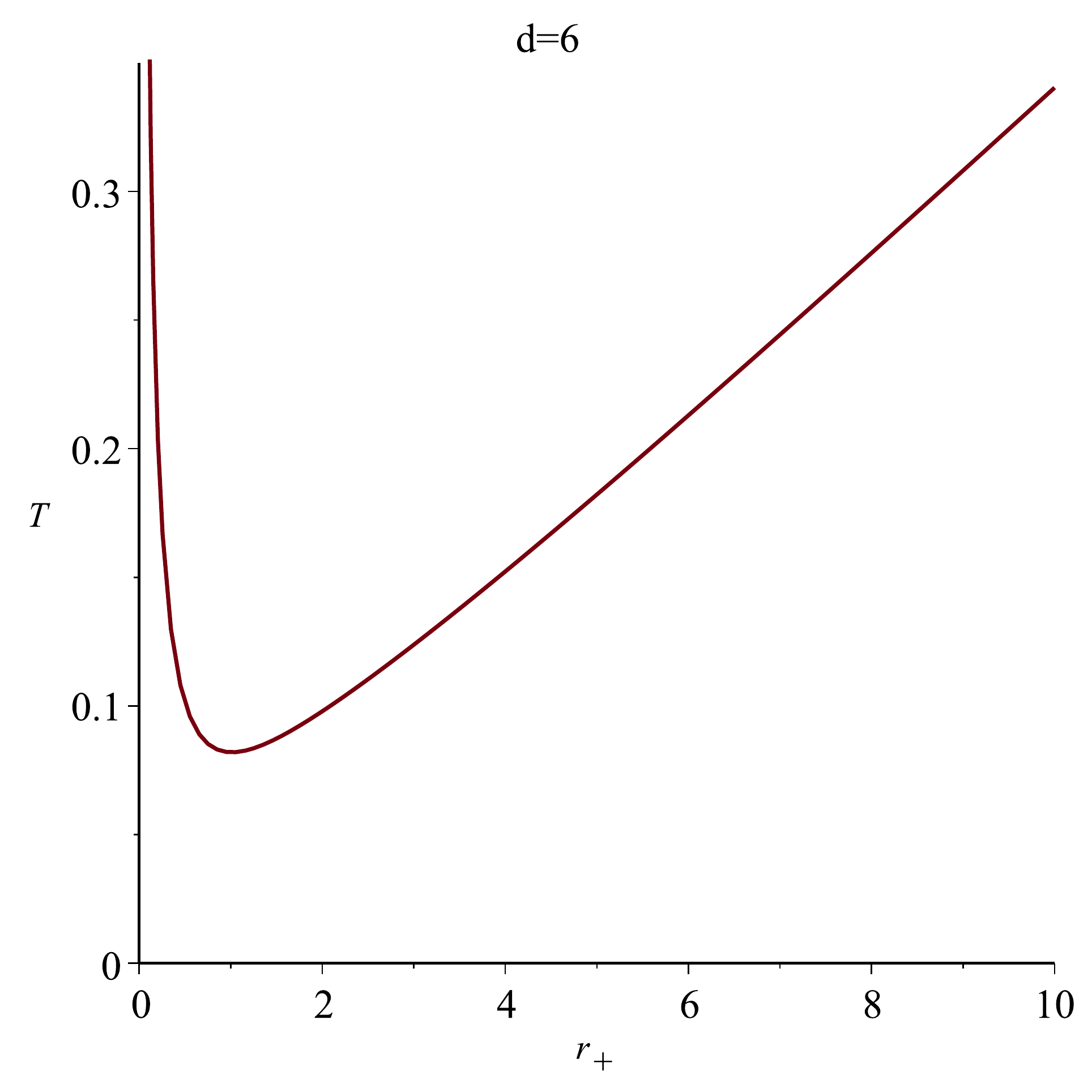}
\vspace{-1mm}
\caption{Behavior of the temperature $T$ as function of the horizon radius $r_+$, corresponding to $d=4$, $d=5$, and $d=6$. We set $\epsilon=0.5$ and $\Lambda_{W}=-1$ in all three cases.}
\label{fig1}
\end{center}
\end{figure}

When $\epsilon=1$, this term vanishes and the system reduces to Einstein gravity, which possesses a Hawking-Page phase transition \cite{Hawking:1982dh}. The temperature $T$ admits a minimum at
\begin{align}
r_c=\sqrt{-\frac{(d-2)}{(d-1)\Lambda_{W}}}(d-3),
\label{7}
\end{align}
and the corresponding temperature is
\begin{align}
T_c=\frac{1}{2\pi}\sqrt{-\frac{d-1}{d-2}\Lambda_{W}}.
\end{align}
However, if $\epsilon\neq1$, the features of the temperature $T$ depend on the spacetime dimensions $d$. In $d=4$ dimension, the temperature
\begin{align}
T=-\frac{3\Lambda_{W}^2r_+^4-2\Lambda_{W}r_+^2+\epsilon^2-1}{8\pi r_+(\Lambda_{W}r_+^2+\epsilon^2-1)}
\end{align}
becomes zero at
\begin{align}
r_+=\sqrt{-\frac{\sqrt{4-3\epsilon^2}-1}{3\Lambda_{W}}}.
\end{align}
In $d=5$ dimension, the term $(1-\epsilon^2)(d-2)^2(d-3)^2(d-5)$ vanishes, and the temperature
\begin{align}
T=-\frac{\Lambda_{W}r_+(\Lambda_{W}r_+^2-3)}{6\pi (\Lambda_{W}r_+^2+3\epsilon^2-3)}
\label{11}
\end{align}
becomes zero at $r_+=0$.
In $d\geqslant 6$ dimensions, the temperature resembles the Schwarzschild-AdS black hole in general relativity and admits a minimal value. In FIG.\ref{fig1} we present some examples on the behavior of the temperature $T$ as function of the horizon radius $r_+$, namely for $d=4$, $d=5$, and $d=6$. We set $\epsilon=0.5$ and $\Lambda_{W}=-1$ in all three cases. For different values of $\epsilon$, there may be critical phenomena in the thermodynamical phase space, but this is not the focus of our study. See \cite{Xu:2018fag} for more discussions.

\section{Black Hole Evaporation in Ho\v{r}ava-Lifshitz Gravity}\label{(III)}

In this section we consider the black hole evaporation process in HL gravity for various spacetime dimension $d$ and detailed balance violation parameter $\epsilon$. Due to the Hawking radiation, the black hole would keep losing mass, so $M$ should be monotonically-decreasing functions of time. For simplicity, we shall assume that the emitted particles are all massless. Applying geometrical optics approximation, all the particles should move along null geodesics. If we orient the extra $(d-3)$ angular coordinates in $\mathrm{d}\Omega_{d-2}^{2}$ and normalize the affine parameter $\lambda$, the geodesic equation of the massless particles reads
\begin{align}
\bigg(\frac{\mathrm{d}r}{\mathrm{d}\lambda}\bigg)^2=E^2-J^2\frac{f(r)}{r^2},
\end{align}
where $E=f(r)\frac{\mathrm{d}t}{\mathrm{d}\lambda}$ and $J=r^2\frac{\mathrm{d}\theta}{\mathrm{d}\lambda}$ correspond to the energy and angular momentum of the emitted particle, respectively. If we consider a null geodesic that is coming from just outside the black hole horizon, it will turn back towards the hole when there is a turning point satisfying $\big(\frac{\mathrm{d}r}{\mathrm{d}\lambda}\big)^2=0$, thus this particle cannot be detected by the observer on the AdS boundary. Defining the impact parameter $b\equiv {J}/{E}$, the massless quanta can reach infinity only if the following condition is satisfied:
\begin{align}
\frac{1}{b^2}> \frac{f(r)}{r^2},
\end{align}
for all $r> r_+$.

The maximal value of ${f(r)}/{r^2}$, which can be used to define the impact factor $b_c$, depends on the exact form of $f(r)$. Once we obtained the impact factor, according to the Stefan-Boltzmann law, we conclude that in $d$-dimensional spacetime, the Hawking emission power is \cite{Vos1989,Cardoso:2005cd,Frolov2011,Xu:2019wak,Xu:2019krv}
\begin{align}
\frac{\mathrm{d} M}{\mathrm{d}t}=-C b_c^{d-2} T^d.
\label{law}
\end{align}
The above formula implies the emission power is proportional to the $(d-2)$-dimensional cross section and the photon energy density in $(d-1)$-dimensional space. The photon energy density in $(d-1)$-dimensional space (spatial dimension only) is proportional to $T^d$, and the cross section of $d$-dimensional non-rotating black hole is $b_c^{d-2}$. Since the $T^d$ term possesses the higher order, the behavior of temperature $T$, especially the asymptotical behavior, is extremely important in black hole evaporation process. We will find the the evaporation process is largely depending on the features of $T$ presented in \eqref{7}-\eqref{11}. The constant $C=(d-2)\pi^{\frac{d}{2}-1}\Omega_{d-2}\frac{\Gamma(d)}{\Gamma(\frac{d}{2})}\zeta(d)$. Since we are only concerned about the qualitative features of the evaporation process, without loss of generality, we set the constant $C=1$. Now let us study the black hole evaporation for various values of $\epsilon$.

\subsection{$\epsilon=0$}
In the $\epsilon=0$ case, the black hole metric reads
\begin{align}
  f(r)=1-\frac{2\Lambda_{W}r^2}{(d-2)(d-3)}
  -r^{\frac{5-d}{2}}\sqrt{-M\Lambda_{W}}.
\end{align}
In large distance approximation, the asymptotically AdS spacetime can be recovered. For convenience, we shall define the ``effective AdS radius'' to be
\begin{align}
\ell\equiv\sqrt{-\frac{(d-3)(d-2)}{2\Lambda_{W}}},
\end{align}
then the black hole metric, mass, and temperature can be written as
\begin{align}
  f(r)=1+\frac{r^2}{\ell^2}-r^{\frac{5-d}{2}}\sqrt{\frac{M(d-3)(d-2)}{2\ell^2}},
\end{align}
\begin{align}
  M=\frac{2r_+^{d-1}}{(d-3)(d-2)\ell^2}+\frac{4r_+^{d-3}}{(d-3)(d-2)}+\frac{2r_+^{d-5}\ell^2}{(d-3)(d-2)},
\end{align}
and
\begin{align}
  T=\frac{(d-1)r_+^2+(d-5)\ell^2}{8\pi r_+ \ell^2}.
\end{align}
For any $d\geqslant 4$ and $d\neq 5$, the function $\frac{f(r)}{r^2}$ admits an extremal point at
\begin{align}
  r^*=\left(\frac{(d-3)(d-2)(d-1)^2M}{32 \ell^2}\right)^{\frac{1}{d-5}}.
\end{align}
For $d=4$, it is a minimum and the ${f(r)}/{r^2}$ reaches maximal value at $r\rightarrow \infty$, while for $d=5$, ${f(r)}/{r^2}$ is a monotonic function of $r_+$, so the impact factor is $b_c=\ell$ in $d=4,5$. For $d\geq 6$, $r^*$ corresponds to the unstable photon orbit, and the impact factor is given by  $b_c= {r^*}/{\sqrt{f(r^*)}}$.

In the case of $d=4$, upon inserting the black hole mass $M$, temperature, and impact factor $\ell$ into Stefan-Boltzmann law, we have
\begin{align}
  \d t=-\frac{4096\pi^4 r_+^2\ell^4(r_+^2+\ell^2)}{(3r_+^2-\ell^2)^3} \d r_+.
\end{align}
The expression on the right hand side becomes divergent at $r_+={\ell}/{\sqrt{3}}$, which corresponds to the $T=0$ state. Integrating the above formula from any initial black hole radius to $r_+={\ell}/{\sqrt{3}}$, we can find the lifetime of the black hole is always divergent. The black hole can lose away a huge amount of mass from arbitrarily large initial mass to a finite mass with in a finite time. However, when the black hole evaporates, the temperature also decreases quickly near $r_+={\ell}/{\sqrt{3}}$, so the evaporation process becomes increasingly difficult. The black hole will take infinite time to completely evaporate away. That is, it effectively becomes a remnant, which may help to ameliorate the information paradox \cite{Chen:2014jwq}. This phenomenon also obeys the third law of black hole thermodynamics.

Similarly, for the case of $d=5$, we have
\begin{align}
  \d t=-\frac{128\pi^5 \ell^5 (r_+^2+\ell^2)}{3r_+^4} \d r_+.
\end{align}
This integral is also divergent near $r_+=0$, which also corresponds to $T=0$.

On the other hand, for the cases $d\geqslant 6$, defining $x\equiv {r_+}/{\ell}$ and $X\equiv \left(x^2+x^{-2}+2\right)x^{d-3}$, we have
\begin{align}
  M=\frac{2X}{(d-3)(d-2)}\ell^{d-3},
\end{align}
\begin{align}
  T=\frac{(d-1)x^2+(d-5)}{8\pi x}\ell^{-1},
\end{align}
and
\begin{align}
  b_c=\frac{\left(\frac{1}{16}X(d-1)^2\right)^{\frac{1}{d-5}}}{\left(\frac{d-5}{d-1}+\left(\frac{1}{16}X(d-1)^2\right)^{\frac{2}{d-5}}\right)^{\frac{1}{2}}}\ell.
\end{align}
Inserting the above three formulas into the Stefan-Boltzmann law, we have
\begin{align}
\mathrm{d}t=\ell^{d-1}F(x,d)\mathrm{d}x,
\end{align}
where $F(x,d)$ is a complicated function which is not worth explicitly written. Integrating the above formula from $\infty$ to 0, we can find that the integral $\int^{0}_{\infty} F(x,d)\mathrm{d}x$ is convergent and it depends only on the value of $d$. Thus, we can conclude that the lifetime of black hole does not diverge with any initial black hole mass, and it is bounded by a time of the order of $\ell^{d-1}$. This result is consistent with the Schwarzschild AdS black hole in Einstein gravity \cite{Page:2015rxa}.

\begin{figure}[h!]
\begin{center}
\includegraphics[width=0.32\textwidth]{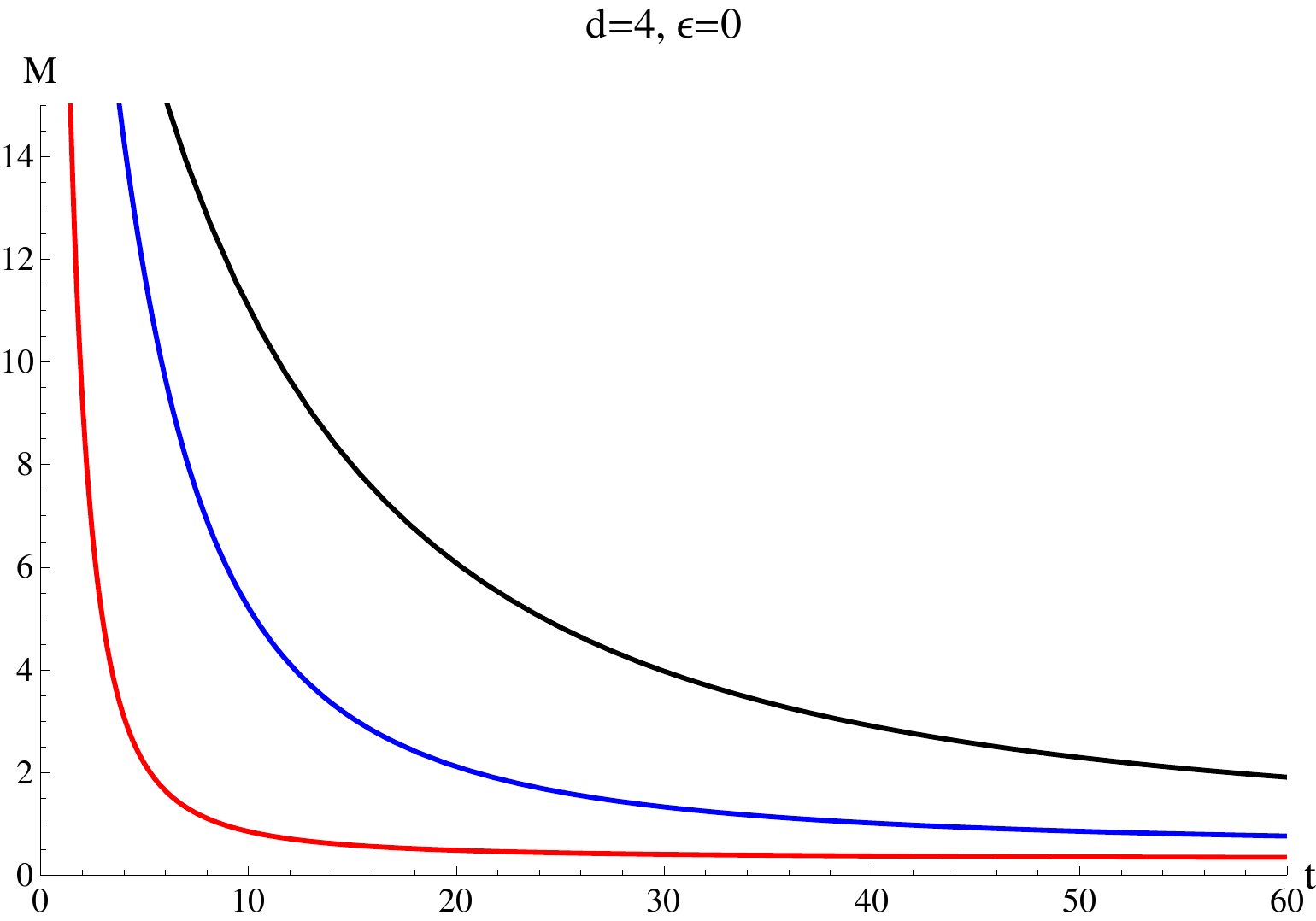}
\includegraphics[width=0.32\textwidth]{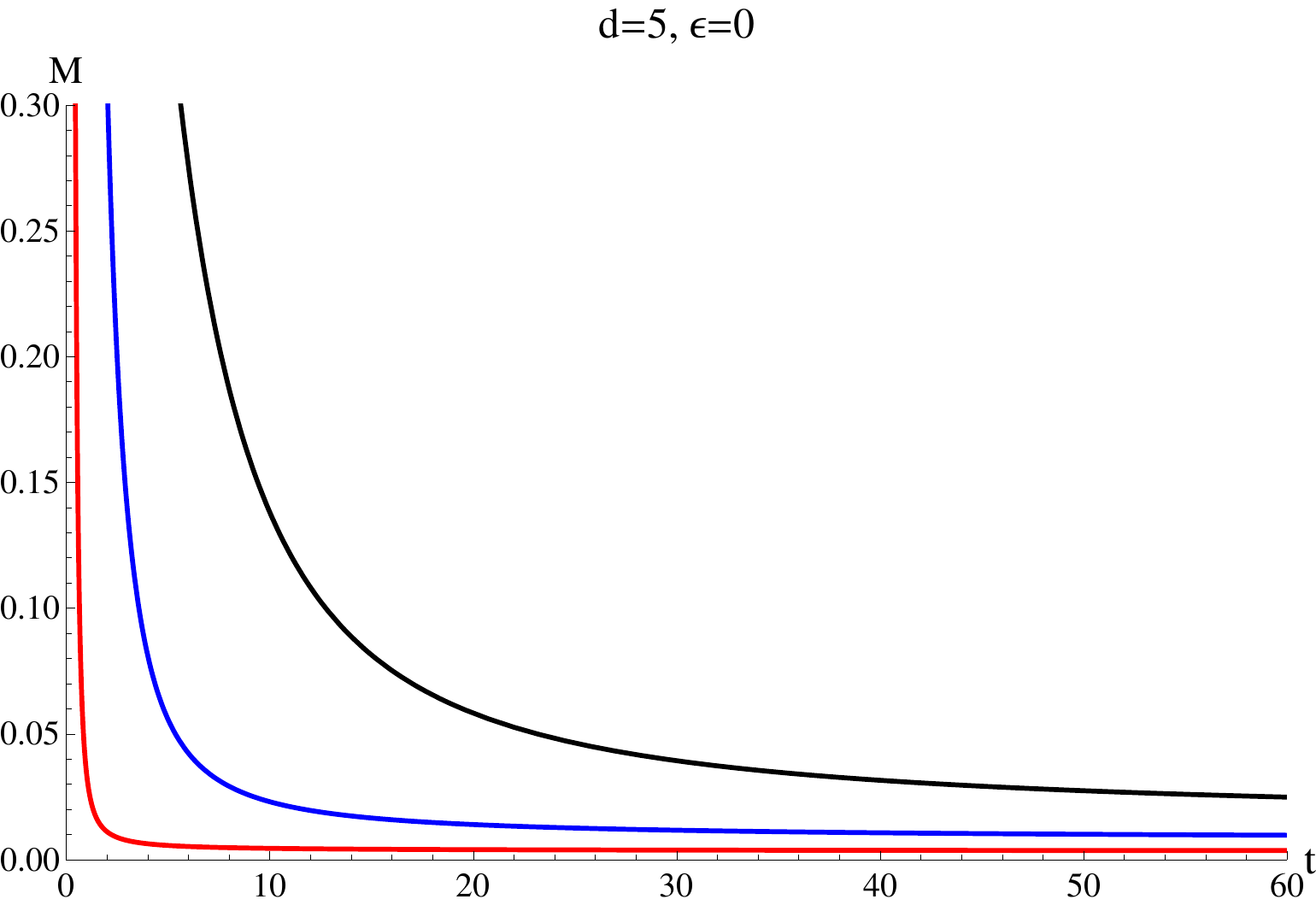}
\includegraphics[width=0.32\textwidth]{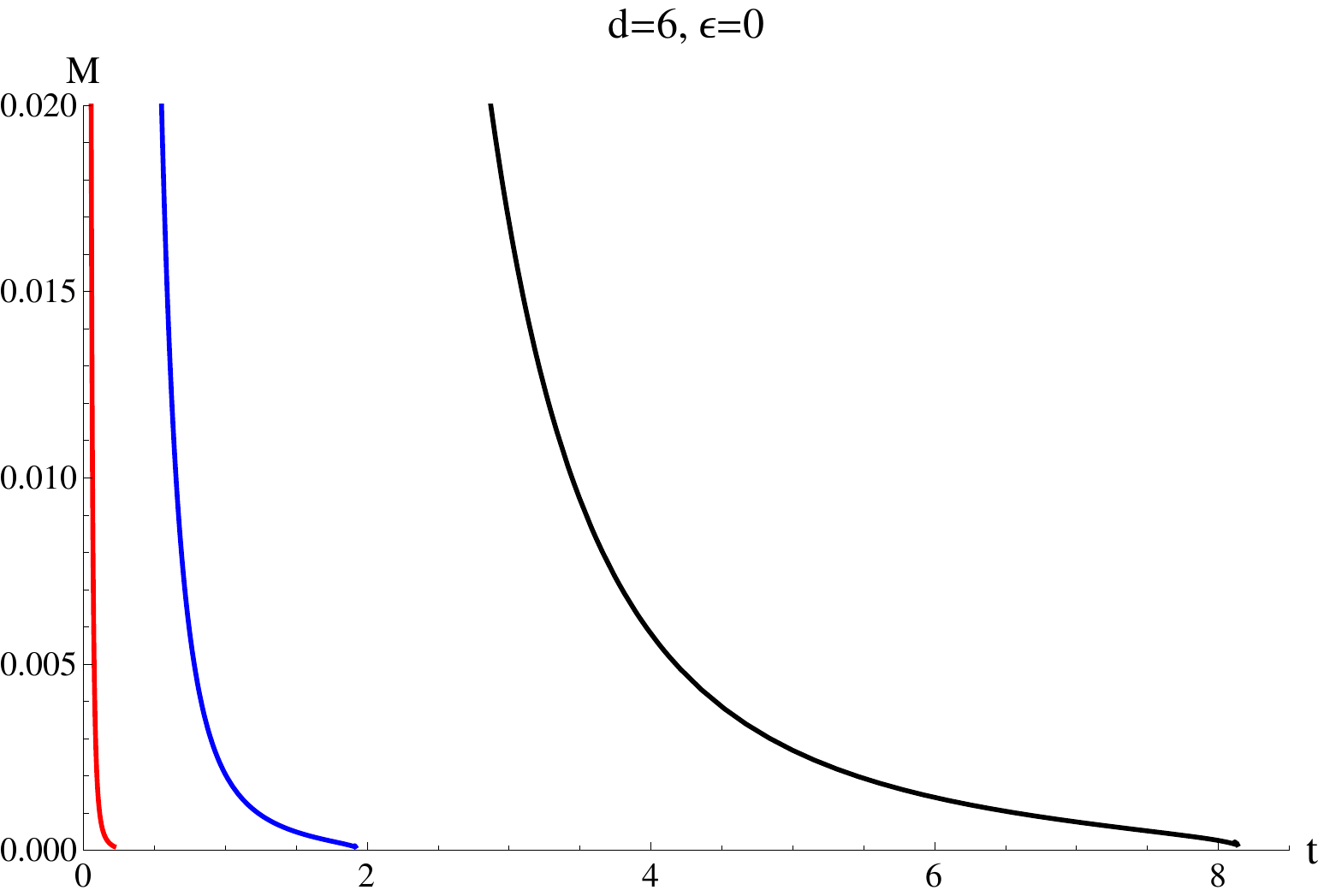}
\vspace{-1mm}
\caption{The evolution of the black hole in $d=4,5,6$ and $\epsilon=0$. The initial mass $M_0$ is taken to infinity. In each figure from left to right the curves correspond to $\ell=0.1$, $\ell=0.15$ and $\ell=0.2$ respectively.}
\label{fig2}
\end{center}
\end{figure}

In FIG.\ref{fig2} we present some numerical examples of the black hole mass as function of time $t$ in $d=4,5,6$ and $\epsilon=0$. The initial mass $M_0$ is taken to infinity. In each figure from left to right the curves correspond to $\ell=0.1$, $\ell=0.15$ and $\ell=0.2$ respectively. In the case of $d=4,5$, the black hole loses its mass quickly at the beginning, but evaporation gets harder when the black hole becomes smaller, thus the black hole will have infinite lifetime. In the case of $d=6$, the black hole can always evaporate away in a finite time. The lifetime of the black hole is in order of $\ell^5$.

\subsection{$\epsilon=1$}
In the case of $\epsilon=1$, the black hole returns to the Schwarzschild AdS solution, with metric function
\begin{align}
  f(r)=1+\frac{r^2}{\ell^2}-\frac{(d-3)(d-2)M}{4r^{d-3}}.
\end{align}
For any $d\geqslant 4$, the function ${f(r)}/{r^2}$ admits an extremal point at
\begin{align}
  r^*=\left(\frac{1}{8}M(d-1)(d-2)(d-3)\right)^{\frac{1}{d-3}},
\end{align}
which corresponds to the unstable photon orbit, thus the impact factor is $b_c= {r^*}/{\sqrt{f(r^*)}}$.
The black hole mass and temperature can be written as, respectively,
\begin{align}
  M=\frac{4r_+^{d-1}}{(d-3)(d-2)\ell^2}+\frac{4r_+^{d-3}}{(d-3)(d-2)},
\end{align}
\begin{align}
  T=\frac{(d-1)r_+^2+(d-3)\ell^2}{4\pi r_+ \ell^2}.
\end{align}
Similarly we can also define $x\equiv {r_+}/{\ell}$ and $\textrm{X}=\left(\frac{d-1}{2}(x^2+1)x^{d-3}\right)^{\frac{1}{d-3}}$, then the black hole mass, temperature, and impact factor can be written as
\begin{align}
  M=\frac{4(x^2+1)(x^{d-3})}{(d-3)(d-2)}\ell^{d-3},
\end{align}
\begin{align}
  T=\frac{(d-1)x^2+(d-3)}{4\pi x}\ell^{-1},
\end{align}
and
\begin{align}
  b_c=\frac{\textrm{X}}{\left(\frac{d-3}{d-1}+\textrm{X}^2\right)^{\frac{1}{2}}}\ell.
\end{align}
Inserting the above formulas into the Stefan-Boltzmann law, we can also obtain the result that the lifetime of the black hole is bounded by a time of the order of $\ell^{d-1}$ \cite{Page:2015rxa,Xu:2019krv}.

\begin{figure}[h!]
\begin{center}
\includegraphics[width=0.32\textwidth]{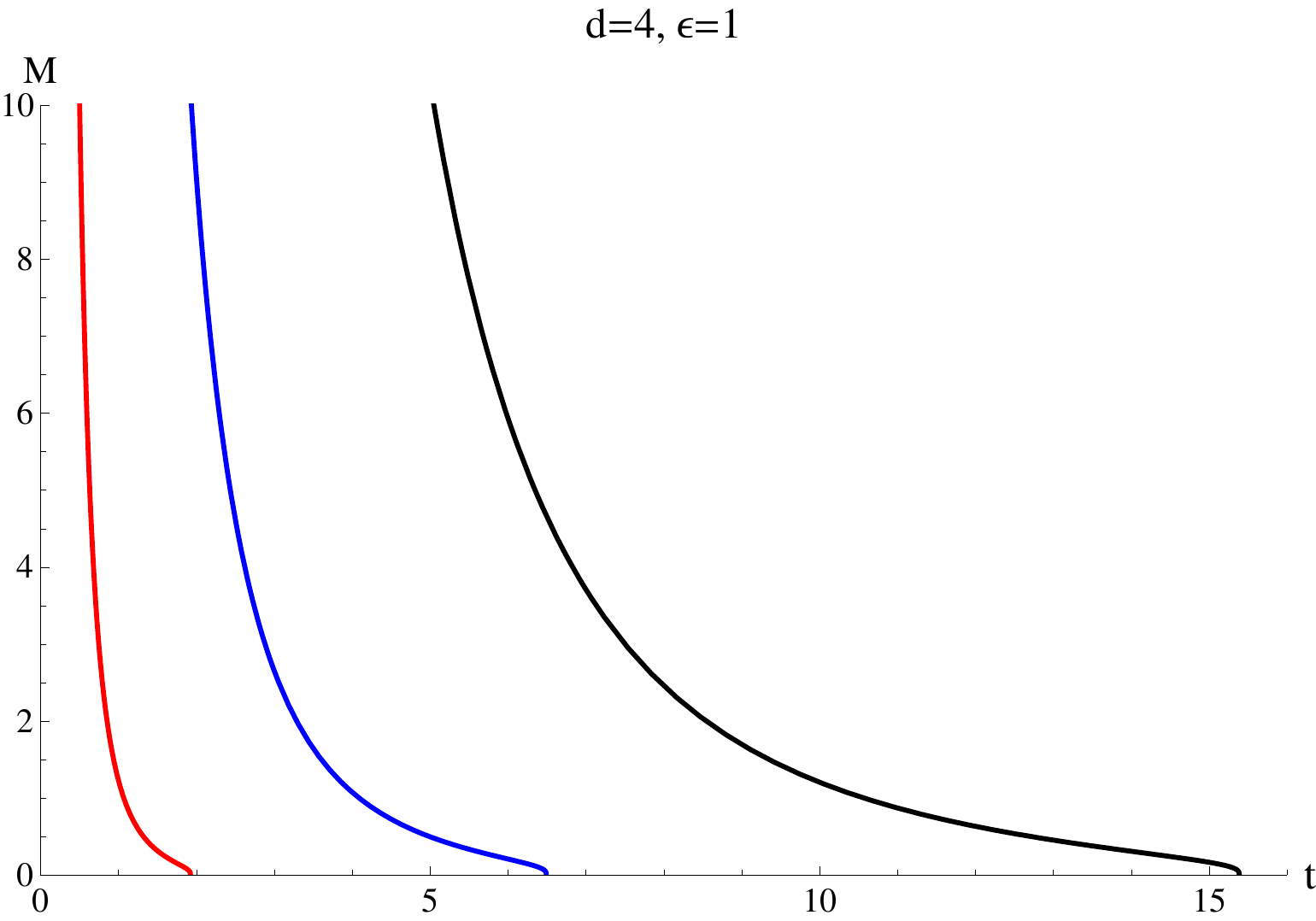}
\includegraphics[width=0.32\textwidth]{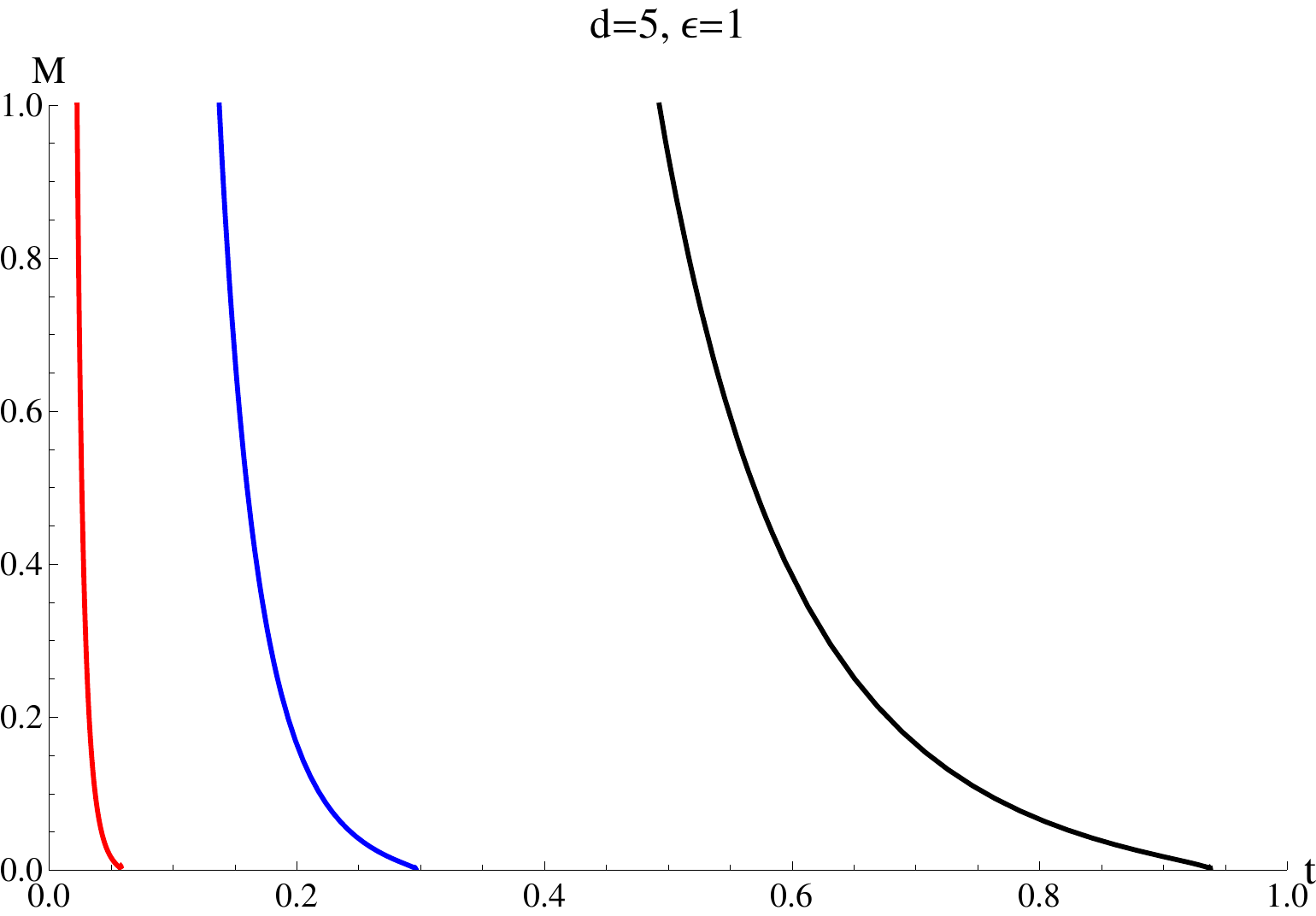}
\includegraphics[width=0.32\textwidth]{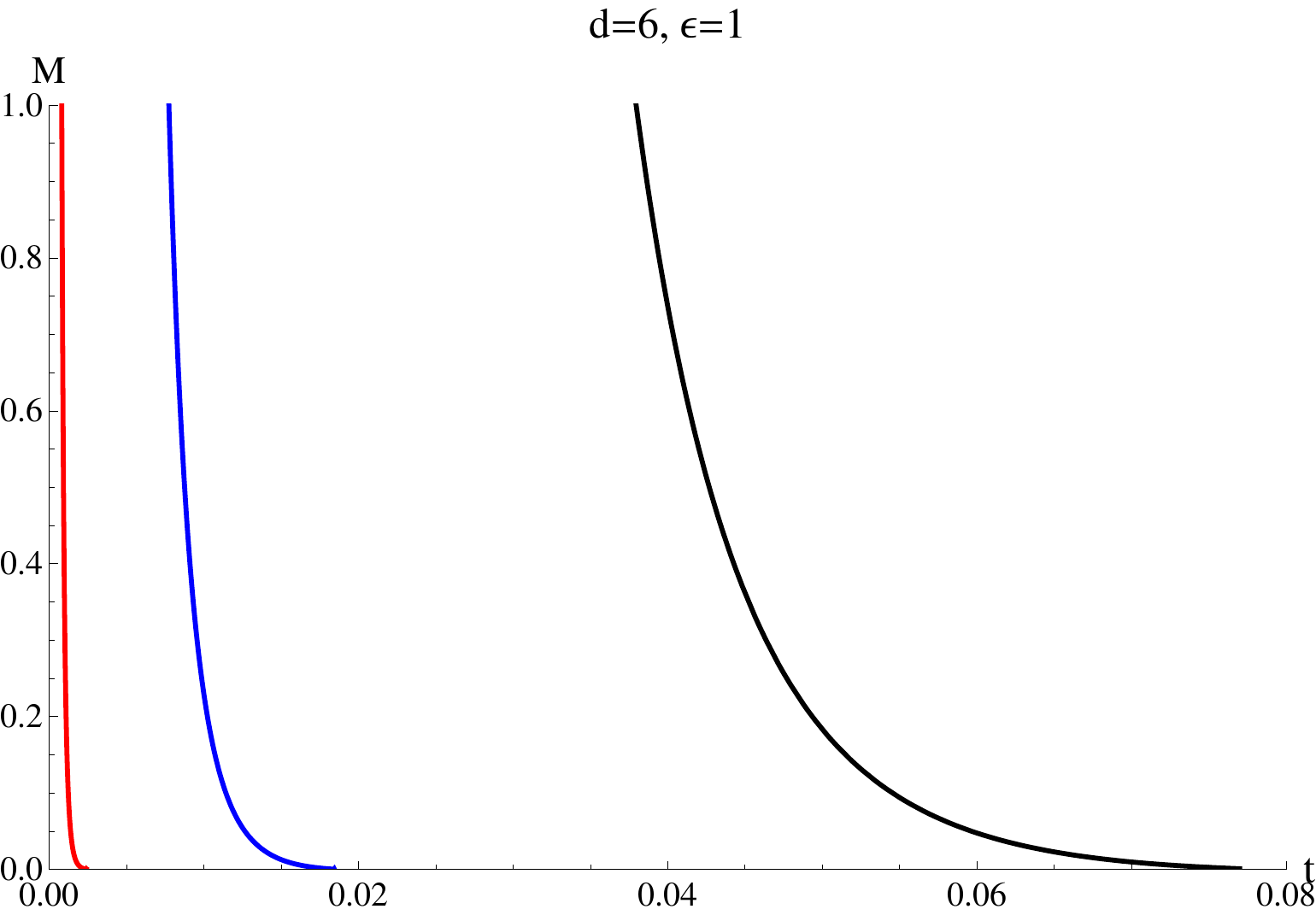}
\vspace{-1mm}
\caption{The evolution of the black hole in $d=4,5,6$ and $\epsilon=1$. The initial mass $M_0$ is taken to infinity. In each figure from left to right the curves correspond to $\ell=0.1$, $\ell=0.15$ and $\ell=0.2$,  respectively.}
\label{fig3}
\end{center}
\end{figure}

In FIG.\ref{fig3} we present some numerical examples of the relationship of black hole mass and time $t$ for in $d=4,5,6$ and $\epsilon=1$. The initial mass $M_0$ is taken to infinity. In each figure from left to right the curves correspond to $\ell=0.1$, $\ell=0.15$ and $\ell=0.2$ respectively. In all three cases, the black holes evaporate away in a finite time. The lifetime of the black hole is in order of $\ell^{d-1}$.

\subsection{$0<\epsilon^2<1$}
In the case of $0<\epsilon^2<1$, the black hole metric takes the form
\begin{align}
  f(r)=1+\frac{r^2}{(1-\epsilon^2)\ell^2}-4r^{\frac{5-d}{2}}\sqrt{\frac{(d-3)(d-2)M}{32(1-\epsilon^2)\ell^2}+\frac{\epsilon^2 r^{d-1}}{16(1-\epsilon^2)^2\ell^4}}.
  \label{metric}
\end{align}
In this situation, if we directly compute $\frac{d}{dr}\frac{f(r)}{r^2}=0$, we can obtain the equation
\begin{align}
64\epsilon^2r^{2d-6}+32M(d-3)(d-2)(1-\epsilon^2)\ell^2r^{d-5}=M^2(d-3)^2(d-2)^2(d-1)^2.
\label{high}
\end{align}
For the cases of $d=4,5$ the above equation can be explicitly solved. For $d=4$, we have
\begin{align}
r^*=\frac{S}{4 \epsilon}+\frac{3M^2}{4S},
\end{align}
where
\begin{align}
S=\left(32\epsilon^3\ell^2M-32\epsilon \ell^2M+M\sqrt{1024\epsilon^6 \ell^4-2018\epsilon^4 \ell^4+1024\epsilon^2 \ell^4-27M^4}\right)^{\frac{1}{3}}.
\end{align}
For $d=5$, we have
\begin{align}
r^*=\frac{(3M\epsilon^2 (3M+\epsilon^2\ell^2-\ell^2))^{\frac{1}{4}}}{\epsilon}.
\end{align}

For higher dimensions, eq.\eqref{high} is a higher power equation, so finding the exact maximal value of ${f(r)}/{r^2}$ is not easy. However, if we consider the large black hole limit $M\rightarrow \infty$, we will find the constant of the second term of lhs of eq.\eqref{high} is much smaller than the rhs, thus it can be omitted. Then we have the photon orbit
\begin{align}
 r^*\approx \left(\frac{M(d-1)(d-2)(d-3)}{8\epsilon}\right)^{\frac{1}{d-3}}.
 \label{r^*}
\end{align}
Since we are interested in the qualitative features of the black hole life, studying the evaporation process of arbitrarily large black holes, the essential question is whether the black hole can lose infinite amount of mass in finite time, and the qualitative features of the black hole evaporation depends on the asymptotical behaviors of the thermodynamical quantities and whether the $T=0$ state exists, so Eq.\eqref{r^*} can still serve as an effective approximation.

\begin{figure}[h!]
\begin{center}
\includegraphics[width=0.32\textwidth]{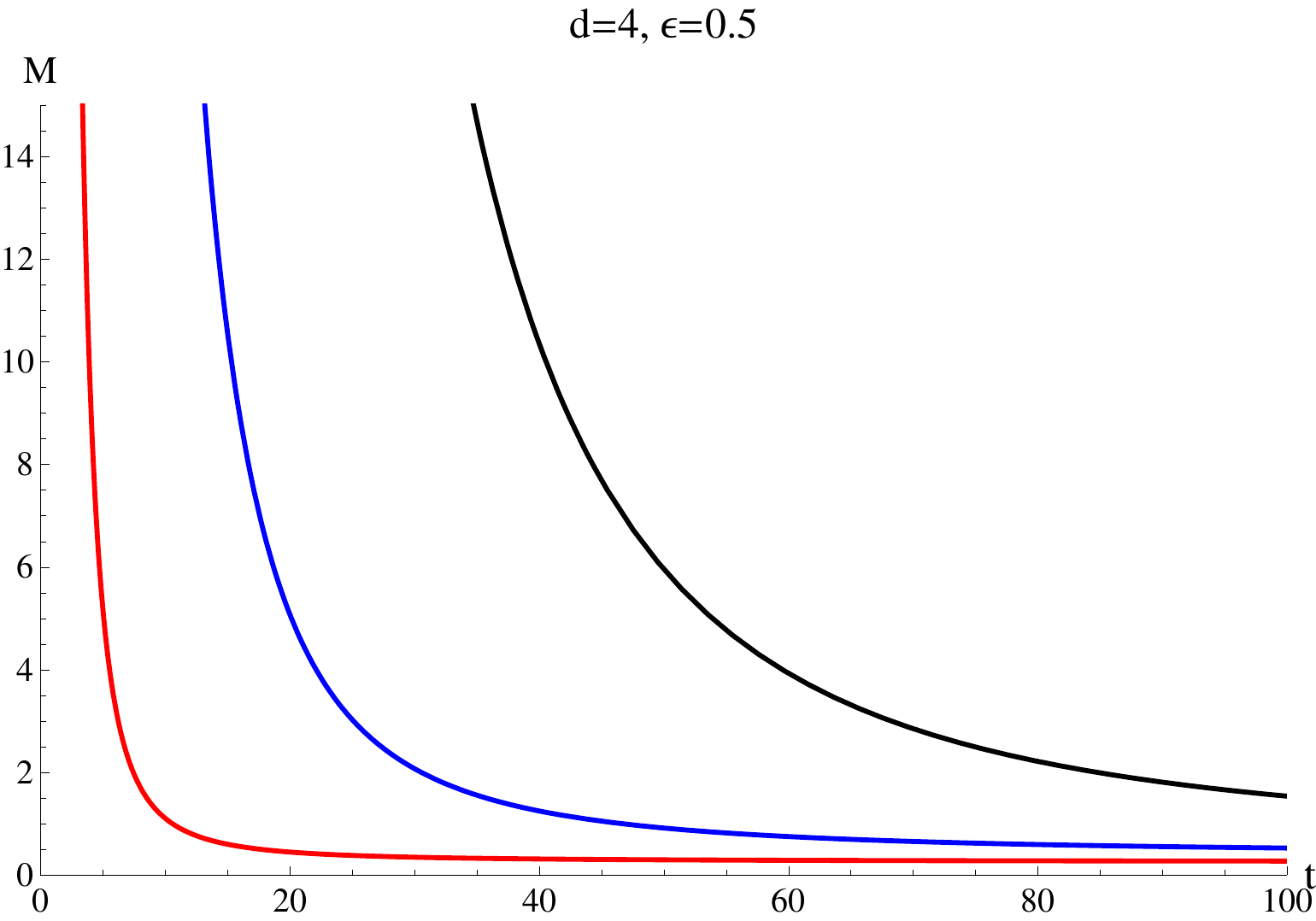}
\includegraphics[width=0.32\textwidth]{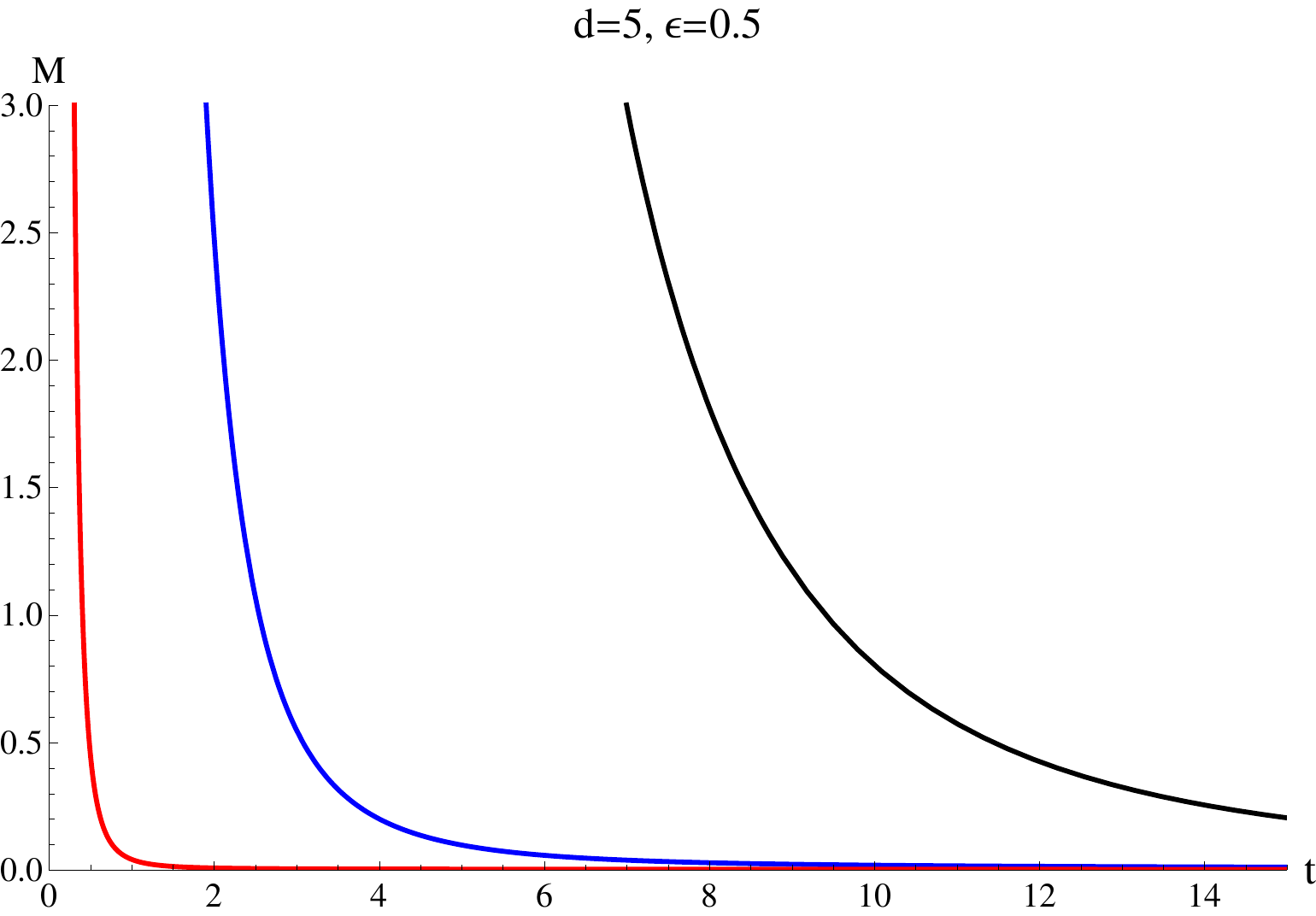}
\includegraphics[width=0.32\textwidth]{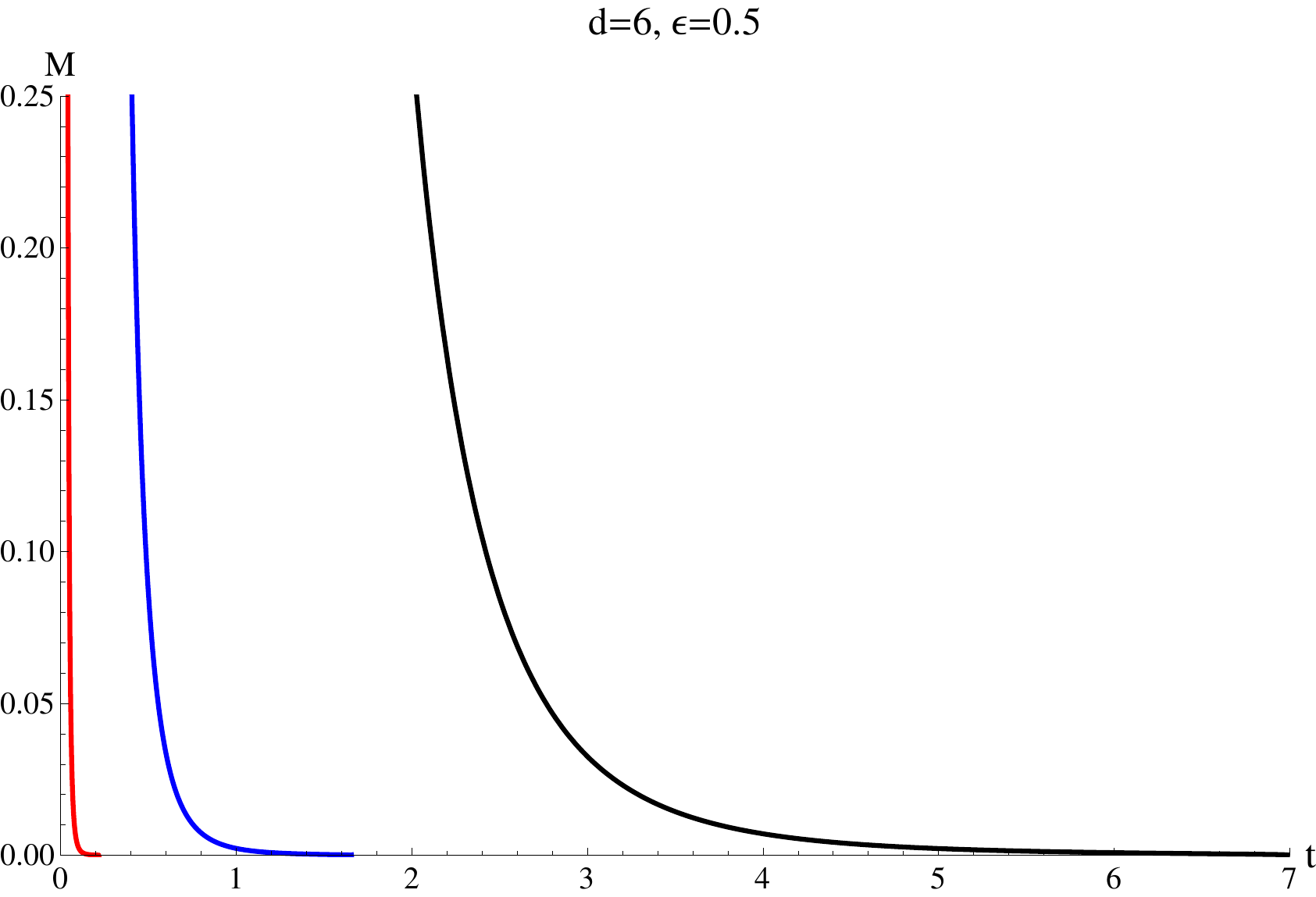}
\vspace{-1mm}
\caption{The evolution of the black hole in $d=4,5,6$ and $\epsilon=0.5$. The initial mass $M_0$ is taken to infinity. In each figure from left to right the curves correspond to $\ell=0.1$, $\ell=0.15$ and $\ell=0.2$,  respectively.}
\label{fig4}
\end{center}
\end{figure}

The black hole mass and temperature are
\begin{align}
  M=\frac{2r_+^{d-1}}{(d-3)(d-2)\ell^2}+\frac{4r_+^{d-3}}{(d-3)(d-2)}+\frac{2 (1-\epsilon^2) r_+^{d-5}\ell^2}{(d-3)(d-2)},
\end{align}
and
\begin{align}
 T=\frac{(d-1)r_+^4+2(d-3)r_+^2\ell^2+(1-\epsilon^2)(d-5)\ell^4}{8\pi r_+\ell^2(r_+^2+(1-\epsilon^2)\ell^2)}.
\end{align}
For the cases of $d=4,5$, since the finial states correspond to $T=0$, according to the third law of thermodynamics, any initial black hole cannot evaporate away in a finite time. In the case of $d\geqslant 6$, using scaling analysis we know we can write the $M$, $T$ and $b_c$ as the functions of $x\equiv {r_+}/{\ell}$ and $\ell$, and we have $M\sim \ell^{d-3}$, $T\sim \ell^{-1}$, and $b_c\sim \ell$. The value of the $\epsilon$ also does not change the qualitative features of these thermodynamical quantities, thus the lifetime of the black hole is still bounded by a time of the order $\ell^{d-1}$. The evaporation process is qualitatively similar to the cases of $\epsilon=0$ in FIG.\ref{fig2}. In FIG.\ref{fig4} we present some numerical results.

\section{Summary}

In order to solve the non-renormalization problem in Einstein gravity and the ghost problem in higher derivative gravity theories, HL gravity chooses to violate the Lorentz invariance. Although Lorentz invariance is one of the fundamental principles of modern physics and is also strongly supported by observations, it seems there are indeed some good reasons to do so. Due to the anisotropy between time and space at short distances, we can have high-order spatial derivative terms in the Lagrangian action, while keeping the time derivative terms to second order, thus the UV behavior is power-counting renormalizable at large energy scale, and the classical Einstein gravity is restored in the IR limit.

In the present work we explore the black hole evaporation process in HL gravity with various spacetime dimensions $d$ and detailed balance violation parameter $0\leqslant \epsilon^2 \leqslant 1$. For the case of $\epsilon=0$, with $d=4,5$, the black hole admits the state of $T=0$, although black hole can lose infinite amount of mass from divergent initial mass to some finite mass, the lifetime of the black hole is always divergent. When the black hole evaporates, the temperature also decreases quickly, so the evaporation process is increasingly difficult. The black hole will take infinite time to completely evaporate away. This phenomenon also obeys the third law of black hole thermodynamics. As $d\geqslant 6$, the lifetime of black hole does not diverge with any initial black hole mass, and it is bounded by a time of the order of $\ell^{d-1}$. The case of $0<\epsilon^2<1$ is also qualitatively similar with $\epsilon=0$. When $\epsilon^2=1$, the black hole solution and thermodynamics reduce to the Schwarzschild-AdS one in Einstein gravity, so the lifetime of black hole is in the order of $\ell^{d-1}$ for any $d\geqslant 4$, which is consistent with the result in \cite{Page:2015rxa}. The black hole evaporation process in $\epsilon\neq 1$ and $d=4,5$ is different from Schwarzschild-AdS because of the existence of the state $T=0$ and the third law of the black hole thermodynamics. The black hole evaporation in HL gravity is different from the other higher derivative gravity theories, such as the conformal (Weyl) gravity \cite{Xu:2017ahm,Xu:2018liy} and the Lovelock gravity \cite{Xu:2019krv}, because the black hole mass in HL gravity can be written as $M\sim\ell^{d-3}$, thus according to the Stefan-Boltzmann law, the lifetime is in the order of $\ell^{d-1}$ in $d\geqslant 6$.

In the present work we only consider the qualitative features of the black hole evaporation, so we omit some coefficients containing $d$, such as the factor $\frac{\Omega_{d-2}}{16\pi(d-3)}$ in $M$, as well as the constant $C=(d-2)\pi^{\frac{d}{2}-1}\Omega_{d-2}\frac{\Gamma(d)}{\Gamma(\frac{d}{2})}\zeta(d)$ in Stefan-Boltzmann law, where $\Gamma(d)$ is the gamma function and $\zeta(d)$ the Riemann zeta function. If we consider these coefficients, we have
\begin{align}
\frac{\mathrm{d} M}{\mathrm{d}t}=-16(d-2)(d-3)\pi^{\frac{d}{2}}\frac{\Gamma(d)}{\Gamma(\frac{d}{2})}\zeta(d) b_c^{d-2} T^d.
\label{law}
\end{align}
In any finite $d\geqslant 4$, the qualitative features remains the same. However, if we consider the large $d\rightarrow \infty$ limit \cite{Emparan:2020inr}, the coefficient on the right hand side of the above formula goes to infinity. We also have $M\sim {1}/{d}$, $T\sim d$, and $b_c\sim\ell$. Furthermore, due to the huge increase in the phase space available to high-frequency quanta at large $d$, the typical energy of Hawking quanta is of the order of $dT$, so the black holes at large $d$ behaves as large quantum radiators and they evaporate away quickly. See more discussion in the large $d$ limit and black hole evaporation in \cite{Emparan:2020inr,Holdt-Sorensen:2019tne}.

Lastly, we suggest some topics that can be further investigated. First of all, our work may be extended to black holes with electric charge. For evaporating charged black hole, we need to consider charged emitted particles come from the Schwinger effect as well as the mass loss due to the Stefan-Boltzmann law. Recently, the $d$-dimensional charged black hole evaporation in Einstein gravity and its relation with cosmic censorship have been investigated in \cite{Xu:2019wak}. It would be interesting to consider the differential equations describing the evolution of charged black hole in HL gravity. In the present work we only consider the gravity in $\lambda=1$, so that Einstein gravity can be recovered in large distance approximation, but of course the case for general $\lambda$ could be interesting, especially its modification to the AdS/CFT correspondence. Since the spacetime is no longer asymptotically AdS in $\lambda\neq1$, a non-AdS/non-CFT correspondence may exist \cite{Aharony:2002up}. Various values of $z$ and different topologies of the black hole are worth investigating. In addition, the grey body factors and different quantum field theories \cite{Klemm:1998bb,Kanti:2004nr} in HL gravity background can also be considered.

\section*{Acknowledgements}

We would like to thank Yong-Hui Qi for useful discussion. YCO thanks the National Natural Science Foundation of China (No.11705162, No.11922508) and the Natural Science Foundation of Jiangsu Province (No.BK20170479) for funding support.


\providecommand{\href}[2]{#2}\begingroup
\footnotesize\itemsep=0pt
\providecommand{\eprint}[2][]{\href{http://arxiv.org/abs/#2}{arXiv:#2}}

\end{document}